\documentclass[a4paper,lnbip]{svmultln}
\usepackage{graphicx}
\usepackage{latexsym}
\usepackage{float}
\usepackage{amssymb}
\usepackage{amsmath}
\usepackage{tabularx}
\usepackage{wrapfig}
\usepackage{multirow}
\usepackage{xcolor,colortbl}
\usepackage{todonotes}
\usepackage{threeparttable}
\usepackage{booktabs}
\usepackage{subfigure}

\graphicspath{{img/}}

\begin{document}
\mainmatter 
 
\title{Investigating Differences between Graphical and Textual Declarative
Process Models\thanks{This research is supported by Austrian Science Fund (FWF):
P26140--N15, P23699--N23. The final publication is available at Springer via
http://dx.doi.org/10.1007/978-3-319-07869-4\_17}}
\titlerunning{Graphical and Textual Declarative Process Models}
\author{Cornelia Haisjackl \and Stefan Zugal}
\authorrunning{Haisjackl et al.}

\institute{University of Innsbruck, Austria\\
\email{{cornelia.haisjackl, stefan.zugal}@uibk.ac.at}
}
\maketitle

\vspace{-0.3cm}
\begin{abstract}
Declarative approaches to business process modeling are regarded as well suited
for highly volatile environments, as they enable a high degree of flexibility.
However, problems in understanding  declarative process models often impede
their adoption. Particularly, a study revealed that aspects that are present in
both imperative and declarative process modeling languages at a graphical
level---while having different semantics---cause considerable troubles. In this
work we investigate whether a notation that does not contain graphical
lookalikes, i.e., a textual notation, can help to avoid this problem. Even
though a textual representation does not suffer from lookalikes, in our
empirical study it performed worse in terms of error rate, duration and mental
effort, as the textual representation forces the reader to mentally merge the
textual information. Likewise, subjects themselves expressed that the graphical
representation is easier to understand.

\keywords{Declarative Process Models, Empirical Research, Mindshift Learning
Theory.}
\end{abstract}

\vspace{-0.8cm}
\section{Introduction}
\label{sec:introduction}
In the context of analyzing and designing information systems, the positive
influence of conceptual modeling on understanding and communication has been
documented~\cite{Mylo98}. For example, \textit{business process models}
(\textit{process models} for short) have been employed in the context of
process--aware information systems, service--oriented architectures and web
services~\cite{ReMe}. Recently, \textit{declarative approaches} have received
increasing attention due to their flexibility with respect to modeling and
execution of processes~\cite{ReWe12}. While imperative process models specify
exactly \textit{how} things must be done, declarative models focus on the logic
that governs the interplay of process actions by describing activities that may
be performed as well as constraints prohibiting undesired behavior. Existing
research has addressed technical issues of declarative process models, such as
maintainability~\cite{ZuPW11}, verification~\cite{Pesi08} and
execution~\cite{Bar+13}. Understandability concerns of declarative models, on
the contrary, have been considered only to a limited extent. So far, a study was
conducted focusing on common strategies and typical pitfalls when system
analysts make sense of declarative process models~\cite{BPMDS2013}. The study
revealed that aspects that are present in both imperative and declarative
process modeling languages at a graphical level---while having different
semantics---cause considerable troubles. To understand these findings, we would
like to refer to the theory of \textit{Mindshift Learning}~\cite{ArHa07}. This
theory postulates that, when learning new modeling languages, concepts that are
similar, but still show subtle differences, are most difficult to learn. In this
work, we investigate whether mindshift learning indeed imposes a burden on
understanding declarative process models by conducting an empirical study,
trying to avoid mindshift learning by using a declarative process modeling
notation based on text. We handed out graphical and textual declarative process
models to subjects and asked them to perform sense--making tasks. Results of
this study indicate that the graphical representation is advantageous
because it gives rise to fewer errors, shorter durations, and less mental
effort. Therefore, even though it might be recommendable to avoid representing
declarative models in a way similar to imperative models, a pure textual
representation does not seem to be the right solution.

The remainder of the paper is structured as follows. Sect.~\ref{sec:background}
gives background information. Then, Sect.~\ref{sec:methodology} describes the
setup of the empirical investigation, whereas Sect.~\ref{sec:performance} deals with its
execution and presents the results. Finally, related work is presented
in Sect.~\ref{sec:related_work}, and Sect.~\ref{sec:summary} concludes the
paper.

\vspace{-0.4cm}
\section{Backgrounds}
\label{sec:background}
Next, we present background information on declarative models
(Sect.~\ref{sec:background_dpm}) and present the concept of mental
effort as a measure for understanding
(Sect.~\ref{sec:background_mental_effort}).

\subsection{Declarative Process Models}
\label{sec:background_dpm}
\begin{figure}[htp]
 \centering
 \includegraphics[width=.85\textwidth]{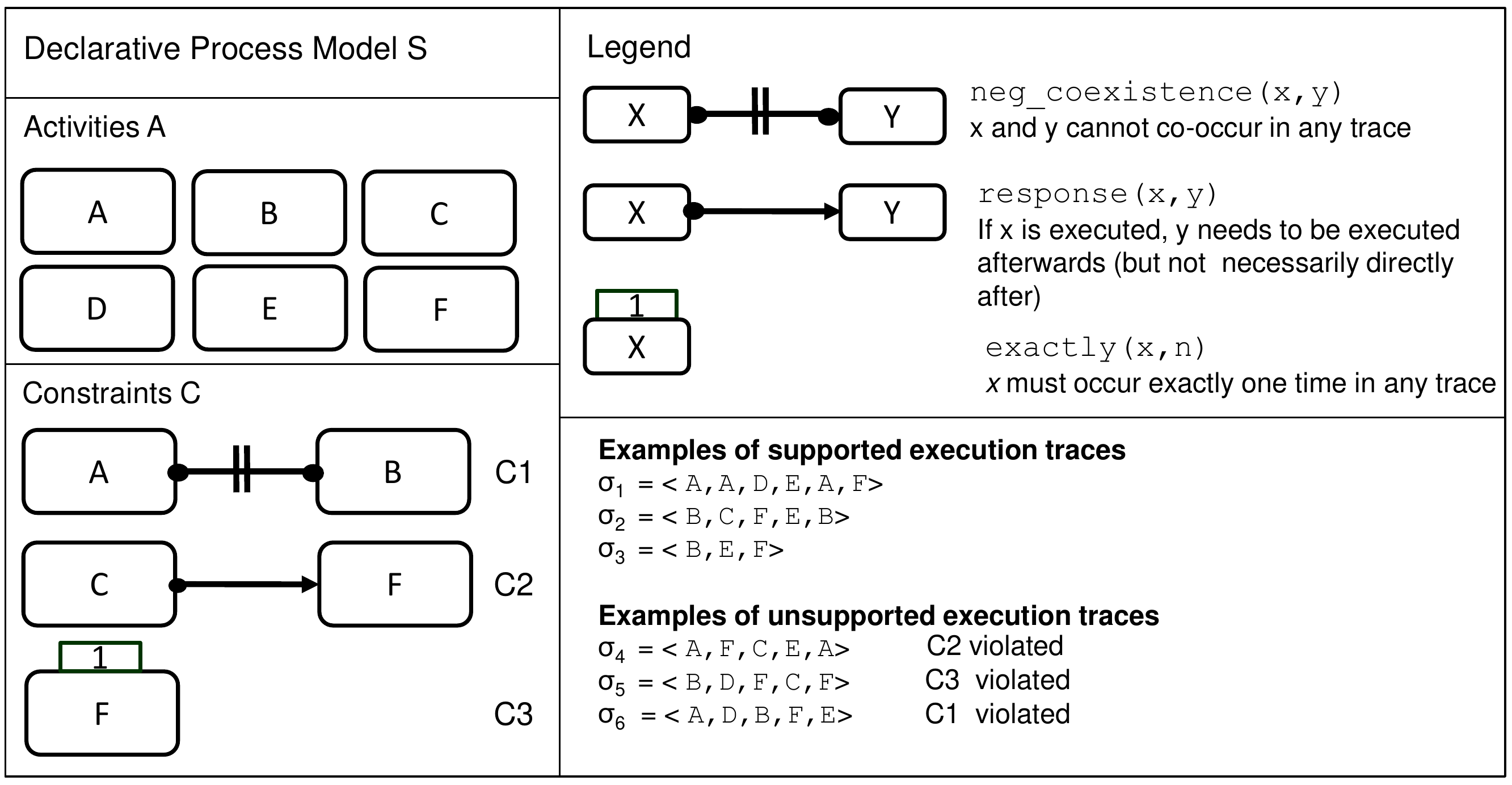}
 \caption{Example of a declarative process model~\cite{ReWe12}}
 \label{fig:example_declModel}
 \vspace{-0.3cm}
\end{figure}

Declarative approaches to business process modeling have received increasing
interest, as they promise to provide a high degree of flexibility~\cite{Pesi08}.
Instead of describing how a process must be executed, declarative models focus on the logic that governs the interplay of
activities. For this purpose, declarative process models specify
\textit{activities} that may be performed as well as \textit{constraints}
prohibiting undesired behavior. Constraints found in literature may be divided
into existence constraints, relation constraints and negation constraints
\cite{AalstPesic2006DecSerFlow}. \textit{Existence constraints} specify how
often an activity must be executed for one particular process instance (e.g.,
$exactly$, cf. \figurename~\ref{fig:example_declModel}). In turn,
\textit{relation constraints} restrict the ordering of activities (e.g.,
$response$, cf. \figurename~\ref{fig:example_declModel}). Finally,
\textit{negation constraints} define negative relations between activities
(e.g., $neg\_coexistence$, cf. \figurename~\ref{fig:example_declModel}).

A \textit{trace} is defined as a \textit{completed} process
instance~\cite{ReWe12}. It can have two different states: either it satisfies
all constraints of the process model (\textit{valid}, also referred to as
\textit{satisfied}), or the trace violates constraints in the process model
(\textit{invalid}, also referred to as \textit{violated}). A \textit{minimal
trace} is defined as a valid trace with a minimum number of activities. A
\textit{sub--trace}, in turn, can be in three different states: First, a
sub--trace can be \textit{valid} (the sub--trace satisfies all constraints of
the process model). Second, it can be \textit{temporarily violated} (the
sub--trace does not satisfy all constraints of the process model, but there is
an affix or suffix that could be added to the sub--trace such that all
constraints are satisfied), or third, \textit{invalid} (the sub--trace violates
constraints in the process model and no affix or suffix can be added to the
sub--trace to satisfy all constraints).

\vspace{-0.4cm}

An example of a declarative process model $S$, using Declare (formerly known as
ConDec)~\cite{Pesi08}, is shown in \figurename~\ref{fig:example_declModel}.
$S$ consists of 6 activities \texttt{A} to \texttt{F} and 3 constraints. The
\textit{neg\_coexistence} constraint (C1) forbids that \texttt{A} and \texttt{B}
co-occur in the same trace. In turn, the \textit{response} constraint (C2)
requires that every execution of \texttt{C} must be followed by one of
\texttt{F} before the process instance may complete. Finally, the
\textit{exactly} constraint (C3) states that \texttt{F} must be executed exactly
once per process instance. For instance, trace \texttt{$\sigma_1$=<A,A,D,E,A,F>}
satisfies all constraints (C1--C3), i.e., these are valid traces, whereas,
e.g., trace  $\sigma_6$ is invalid as it violates C1. Trace
\texttt{$\sigma_7$=<F>} is the minimal trace since there exists no other valid
trace comprising a lower number of activities.

\vspace{-0.4cm}
\begin{figure}[htp]
 \centering
 \includegraphics[width=.85\textwidth]{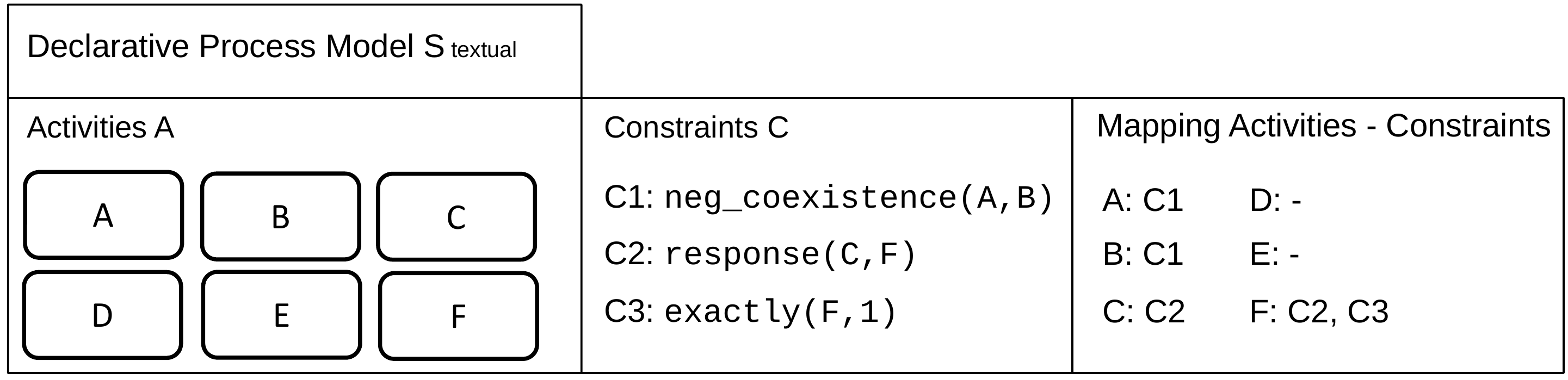}
 \caption{Example of a textual declarative process model}
 \label{fig:example_declModel_text}
 \vspace{-0.3cm}
\end{figure}

In the empirical investigation we try to avoid mindshift learning by using a
declarative process modeling notation based on text.
\figurename~\ref{fig:example_declModel_text} shows the textual representation of
the declarative process model $S$ (cf.
\figurename~\ref{fig:example_declModel}). The textual representation consists of
three parts.  First, a list of activities (activities \texttt{A} to \texttt{F}).
Second, a list of constraints (C1 to C3). Third, an activity--constraint mapping
list, to support subjects when looking up all constraints that are related to a
specific activity (e.g., \texttt{F} is related to constraints C2 and C3).

\subsection{Mental Effort}
\label{sec:background_mental_effort}
To investigate the sense--making of declarative process models, it seems
necessary to also take into account the humans cognitive system---in particular
\emph{working memory}, which is responsible for maintaining and manipulating a
\emph{limited} amount of information for goal--directed behavior, such as the
interpretation of a declarative process model (cf.~\cite{Badd12}). The
amount of working memory currently used is thereby referred to as \textit{mental
effort}~\cite{Paa+03}. Research indicates that a high mental effort increases
the probability of errors, especially when the working memory capacity is
exceeded~\cite{Swel88}. In the context of conceptual models, \cite{Zug+12b}
argues that higher mental effort is in general associated with lower
understanding of models. Various techniques exist for assessing mental effort,
including pupillometry, heart--rate variability and rating scales~\cite{Paa+03}.
Especially rating scales, i.e., self--rating mental effort, has been shown to
reliably measure mental effort and is thus widely adopted~\cite{Paa+03}.
Furthermore, this kind of measurement can be easily applied, e.g., by using
7--point rating scales. In the context of conceptual modeling, it was argued
that mental effort should be considered as an additional measure of understanding
together with error rates and duration~\cite{Zug+12b}.

\vspace{-0.4cm}
\section{Defining and Planning the Empirical Investigation}
\label{sec:methodology}
To investigate whether mindshift learning indeed imposes a burden on
understanding declarative process models we conduct an empirical investigation.

\subsubsection{Research Question.} 
Goal of this empirical investigation is to avoid difficulties because of
mindshift learning due to similarities between imperative and declarative
modeling notations. Therefore, we investigate how system analysts answer several
tasks about declarative process models, once with a graphical model
representation (with presence of mindshift learning) and once with a textual
model representation (with absence of mindshift learning). In particular, we are
interested in differences between graphical and textual model representations
regarding errors, duration and mental effort. Therefore, our research
questions can be stated as follows:\\

\noindent \textbf{Research Question {\boldmath$RQ_{1.1}$}} \textit{What are the
differences between a graphical and textual representation regarding error
rates?}\\
\noindent \textbf{Research Question {\boldmath$RQ_{1.2}$}} \textit{What are the
differences between a graphical and textual representation regarding
duration?}\\
\noindent \textbf{Research Question {\boldmath$RQ_{1.3}$}} \textit{What are the
differences between a graphical and textual representation regarding mental
effort?}\\

With our last research question, we take a broader perspective and ask subjects
directly for advantages and disadvantages for each representation as well as
personal suggestions for improving the understandability of declarative process
models.\\

\noindent \textbf{Research Question {\boldmath$RQ_{2}$}} \textit{What are
advantages of each representation and what are potential improvements for the
understandability of declarative process models?}

\subsubsection{Subjects.} To ensure that obtained results are
not influenced by unfamiliarity with declarative process modeling, subjects need
to be sufficiently trained. Even though we do not require experts, subjects should
have at least a moderate understanding of declarative processes' principles.

\subsubsection{Objects.} 
\begin{figure}[htb]
    \centering
    \subfigure[$P_{1\_graphical}$]{
    \includegraphics[width=.48\textwidth]{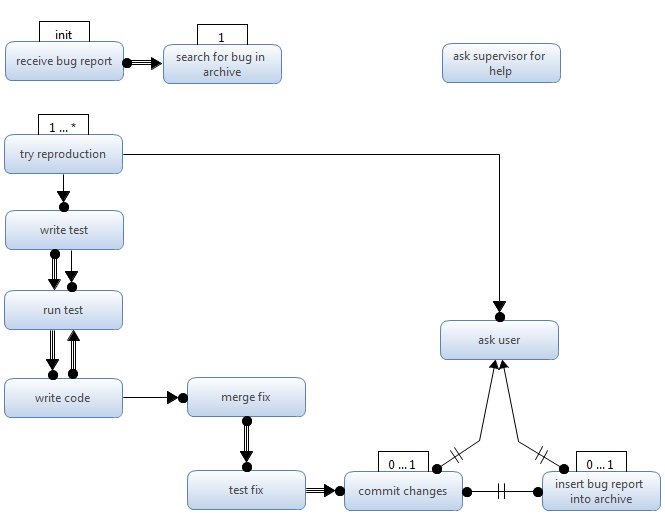}
    \label{fig:P1_graphical_small}}
    \subfigure[$P_{1\_textual}$]{%
    \includegraphics[width=.48\textwidth]{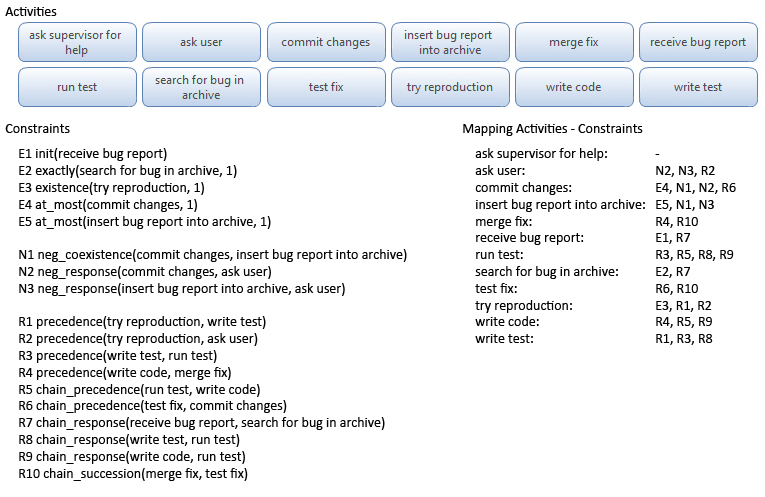}
    \label{fig:P1_textual_small}}
    \caption{Graphical and textual variant of $P_1$}
    \label{fig:ProcessModels}
\vspace{-0.4cm}
\end{figure}

The process models ($P_1$ and $P_2$) used in this
investigation originate from a previous study (cf.~\cite{BPMDS2013}) and
describe real--world business processes. Since we were interested in the
influence of differences regarding the process models' representation, we created a second
variant of each process model describing the exact same process, but with a
textual representation. The variants for $P_1$ are illustrated in
\figurename~\ref{fig:ProcessModels}. For the graphical models we
use the declarative process modeling language Declare~\cite{Pesi08}, where
activities are represented as boxes and constraints as lines or arcs. The
textual models are described in Sect.~\ref{sec:background_dpm}.\footnote{The
empirical investigation's material can be downloaded
from:\\http://bpm.q-e.at/GraphicalTextualDPM}


\vspace{-0.4cm}


The models vary regarding the number of activities (between 12 and 24), number
of constraints (between 18 and 25) and degree of interconnectivity of
constraints, i.e., models consist of 3 to 6 components (cf.~\cite{BPMDS2013}).
The process models are based on two different domains describing bug fixing in a
software company and a worker's duties at an electronic company. Both models
contain constraints of all three types, i.e., existence, relation and negation
constraints.


\subsubsection{Design.} 
\begin{figure}
\begin{center}
  \includegraphics[width=\textwidth]{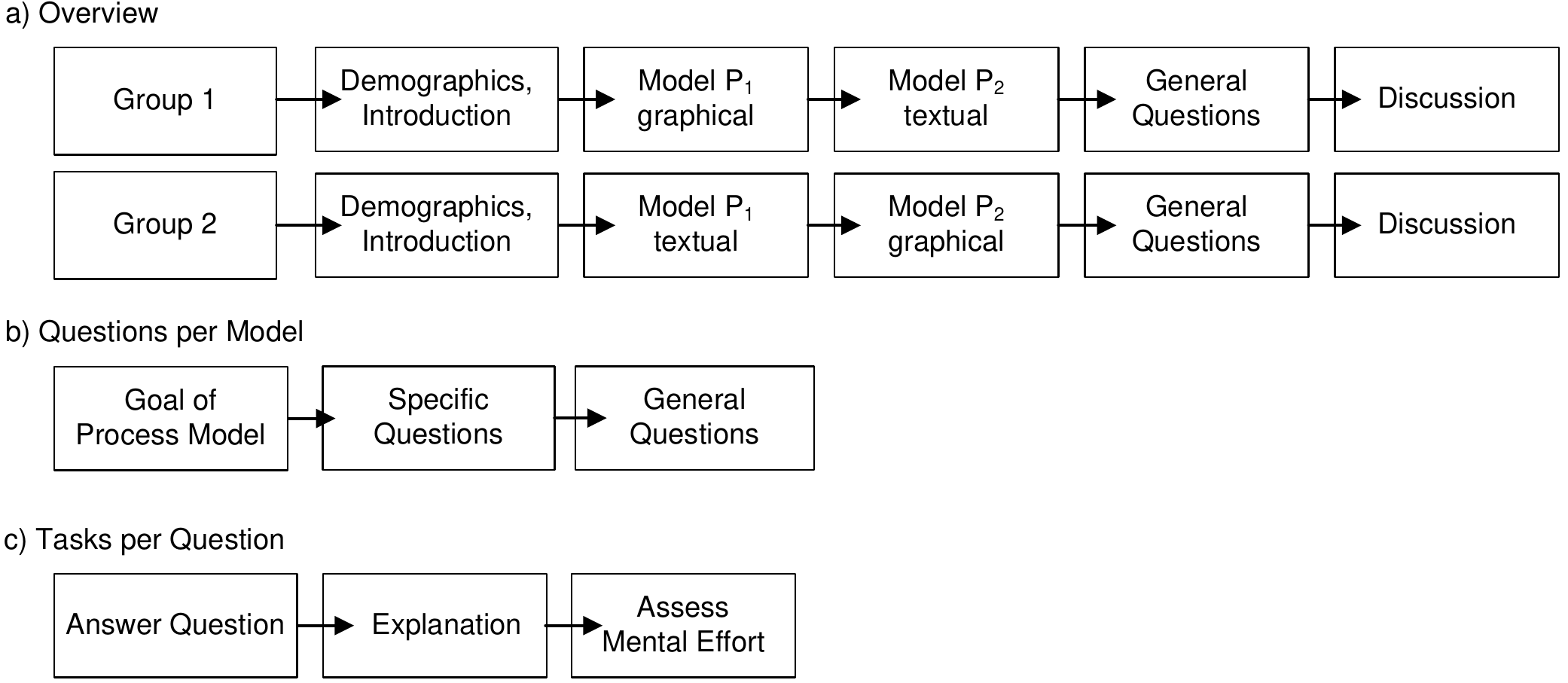}
  \caption{Design of the empirical investigation}
  \label{fig:experimental_design}
  \vspace{-0.7cm}
\end{center}
\end{figure}

\figurename~\ref{fig:experimental_design} shows the
overall design of the empirical investigation: First, subjects are
\textit{randomly} assigned to two groups of similar size. Regardless of the
group assignment, demographical data is collected and subjects obtain
introductory assignments. To support subjects, sheets briefly summarizing the
constraints' semantics are provided, which can be used throughout the
investigation. Then, each subject works on one graphical and one textual process
model. Group 1 starts with the graphical representation of $P_1$, while Group 2
works on the textual representation of the same model. A session is concluded by
a discussion with the subject to help reflecting on the investigation and
providing us with feedback. For each process model, a series of questions is
asked (cf. \figurename~\ref{fig:experimental_design}b): First, subjects are
asked to describe what the goal of the process model is, allowing subjects to
familiarize with the model. Second, we seek to assess whether subjects
understand the process model by asking 3 questions regarding traces in
declarative process models: naming the minimal trace, naming 2 valid traces and
naming 2 invalid traces (cf. Sect.~\ref{sec:background_dpm}). Further, a
series of questions is designed based on the findings of~\cite{BPMDS2013} to
investigate hidden dependencies, pairs of constraints, combinations of
constraints and existence constraints. Third, we ask the subjects about their
opinion on advantages and disadvantages of each model representation, what parts
are most challenging and if they have any suggestions to make the model easier
to read/understand.\\

\vspace{-0.6cm}

For each question, a three--step procedure is followed, cf.
\figurename~\ref{fig:experimental_design} c). First, the subject is asked to
answer the question either by \textit{True}, \textit{False} or \textit{Don't
Know}. Second, the subject has to fill in an explanation field, where it
should be reasoned why the specific answer was given. Third, the subject is
asked to assess the expended mental effort. To this end, a 7--point rating scale
is used, which is known to reliably measure mental effort~\cite{Paa+03}.

\subsubsection {Instrumentation.} For the operationalization of this setup, we
relied on Cheetah Experimental Platform (CEP)~\cite{PiZW10}.
CEP guided the subjects through the sessions, starting with an initial
questionnaire, two questionnaires about declarative process models (one
represented graphically and one textually), a concluding questionnaire and a
feedback questionnaire. Data was collected automatically, ensuring that each
session, the collected demographic data was stored as a separate case of the
empirical investigation.

\vspace{-0.4cm}
\section{Execution and Results}
\label{sec:performance}
So far we described the experimental setup, next we briefly describe its
execution as well as the results.

\subsubsection{Execution} The empirical investigation was conducted in December
2013 at the University of Innsbruck in the course of a weekly lecture on
business processes and workflows; all in all 9 students participated. To prepare
the students, a lecture on declarative process models was held one week before
the empirical investigation. In addition, students had to work on several
modeling assignments using declarative processes before the investigation took
place. Immediately before the sessions, a short lecture revisiting the most
important concepts of declarative process models and the setup was held. The
rest of the session was guided by CEP's experimental workflow
engine~\cite{PiZW10}, as described in Sect.~\ref{sec:methodology}.

\subsubsection{Data Validation} Since our research setup requires subjects to be at least
moderately familiar with Declare, we used a Likert scale, ranging
from \textit{``Strongly agree'' (7)} over \textit{``Neutral'' (4)} to
\textit{``Strongly disagree'' (1)} to screen for familiarity with Declare.
The computed mean is 4.11 (slightly above
average). For confidence in understanding Declare models a mean value of 4.11
was reached (slightly above average). Finally, for perceived competence in
creating Declare models, a mean value of 4 (average) could be computed. Since
all values range about average, we conclude that the participating subjects fit
the targeted profile. In the following, we use the gathered data to investigate
the research questions. 

\subsubsection{$RQ_{1.1}$: What are the differences between a graphical and
textual representation regarding error rates?}
\label{sec:rq1_1}
To investigate $RQ_{1.1}$, the subjects were asked to answer specific questions
(cf. Sect.~\ref{sec:methodology}) As detailed previously, they had to identify one
minimal trace, 2 valid traces and 2 invalid traces for each model.
Since 9 subjects participated in the investigation and each subject worked on
two process models, 18 answers were collected regarding the minimal trace (9 for
each model). Further, 36 for valid traces (18 for each model) and 36 invalid
traces (18 for each model) were collected. Additionally, we asked subjects 2
questions regarding 4 categories for each model. As described in
Sect.~\ref{sec:methodology}, the categories are hidden dependencies, pairs of
constraints, combinations of constraints and existence constraints. Therefore,
there are 9 subjects, 8 questions per model, 2 models, resulting in 144 answers.
Table~\ref{tab:errors} shows the distribution of answers: Overall, subjects gave
179 out of 234 correct answers (76.50\%).

\newcolumntype{f}{>{\raggedleft \arraybackslash} p{1cm}}
\newcolumntype{e}{>{\raggedleft \arraybackslash} p{1.6cm}}
\begin{table}[htb]
        \centering
        \begin{tabular}{lfffeff}
        \toprule
        \multicolumn{1}{c}{}
        &\multicolumn{3}{c}{\textbf{Graphical}}
        &\multicolumn{3}{c}{\textbf{Textual}}
        \\ 
        \textbf{Category}&\textbf{$P_1$}&\textbf{$P_2$}&\textbf{Both}&
        \textbf{$P_1$}& \textbf{$P_2$}&\textbf{Both}\\
       \midrule
       Traces                     & 80\% & 80\% & 80\% &  92\% &
       60\% & 78\% \\
       Hidden Dependencies        &100\% & 90\% & 94\% & 90\% &
       75\% & 83\%\\
       Pairs of Constraints        & 75\% & 60\% & 67\% & 70\% &
       63\% & 67\%\\
       Combination of Constraints & 63\% & 90\% & 78\% & 70\% &
       50\% & 61\%\\
       Existence Constraints      & 63\% & 90\% & 78\% & 70\% &
       75\% & 72\%\\
       \midrule
       Overall                    & 77\% & 82\% & 79\% & 82\% &
       63\% & 74\% \\ \bottomrule
       \end{tabular} 
    \caption{Percentage of correct answers}
    \label{tab:errors}
\end{table}

As mentioned in Sec.~\ref{sec:methodology}, we asked subjects to give us an
explanation for each answer. We used these explanations for identifying and
classifying reasons for errors. Table~\ref{tab:errorsQual} gives an overview of
the data analysis: Overall, 55 answers were incorrect (23.50\%).

\begin{table}[htb]
        \centering
        \begin{tabular}{lccc}
        \toprule
        \textbf{Category}&\textbf{Graphical}&\textbf{Textual}&\textbf{Both}\\
       \midrule
       Subtrace definition        & 8 & 7 & 15 \\
       Overlooked model elements  & 4 & 6 & 10 \\
       Unknown                    & 5 & 5 & 10 \\
       Constraint definition      & 3 & 6 & 9 \\
       Lacking modeling knowledge & 3 & 2 & 5 \\
       Hidden dependency          & 1 & 3 & 4 \\
       Problem with setup         & 0 & 2 & 2 \\
       \bottomrule
       \end{tabular} 
    \caption{Error analysis}
    \label{tab:errorsQual}
    \vspace{-0.6cm}
\end{table}

All in all, we could identify 7 categories why subjects failed to give a correct
answer. Considering the most commonly reason for errors, 15 times subjects
answered incorrectly because they had problems with the definition of a
sub--trace (cf. Sect.~\ref{sec:background_dpm}). Ten times a wrong answer was
given due to overlooked model elements, i.e., activities or constraints.
Additionally 10 times we were not able to categorize the error, because either
the subject did not enter an explanation or the explanation was not sufficient.
Nine times the subjects answered incorrectly due to problems with constraint
definitions, e.g., confusing two constraints with each other. Five errors were
caused by lack of modeling knowledge. Four times a wrong answer was given due to
hidden dependencies. Two times we identified that an error was made because of a
problem with the setup, i.e., we asked for two valid traces, but the subjects
just entered one.

Overall, 31 out of 55 error are due to problems with the setup (either
direct problems with setup or indirect, i.e., lack of knowledge or troubles with
definitions) and 10 unknown. The 14 remaining errors were made
because of overlooking model elements when combining constraints, or hidden
dependencies (cf. \cite{BPMDS2013}). 

\paragraph{Discussion.} In general, we observed that subjects make less errors
when the model is represented graphically. As previous findings~\cite{BPMDS2013}
indicate that subjects have considerable problems making sense of graphically
represented pairs of constraints, we expected that subjects would give fewer
wrong answers using the textual representation. However, our findings indicate that
there is no difference between textual or graphical representation in this
category. It seems that having the disadvantage of mindshift learning is still
less challenging for subjects than the extraction of information from text,
i.e., information that needs to be computed in the human mind~\cite{ScRo96}.

\subsubsection{$RQ_{1.2}$: What are the differences between a graphical and
textual representation regarding duration?}
\label{sec:rq1_2}
To target this research question, we investigated how long it took subjects to
answer all specific questions (c.f., Section~\ref{sec:methodology}).
Table~\ref{tab:durations} shows the duration in minutes for the 11 questions per
model.

\vspace{-0.3cm}
\begin{table}[htb]
        \centering
        \begin{tabular}{lccc}
        \toprule
        &\textbf{Minimum}&\textbf{Maximum}&\textbf{Mean}\\
        \midrule
        $P_1$ graphical representation & 17 & 41 & 28 \\
        $P_1$ textual representation   & 23 & 55 & 37 \\\midrule
        $P_2$ graphical representation & 10 & 20 & 15 \\
        $P_2$ textual representation   & 19 & 30 & 24 \\
        \bottomrule
        \end{tabular} 
    \caption{Duration in minutes}
    \label{tab:durations}
    \vspace{-0.8cm}
\end{table}


\paragraph{Discussion.}
The findings obtained in $RQ_{1.2}$ indicate that answering questions about a
graphically represented model needs less time than for a textual model. In
particular, the disadvantage of mindshift learning is not only less challenging
for subjects than the extraction of information from text (c.f., $RQ_{1.1}$),
but it also needs less time.

\subsubsection{$RQ_{1.3}$: What are the differences between a graphical and
textual representation regarding mental effort?}
\label{sec:rq1_3}
When investigating the sense--making of declarative process models, it seems
desirable to have measures that allow researchers to assess in how far
proposed concepts support the human mind in interpreting declarative process
models. As described in Sect.~\ref{sec:background_mental_effort}, the
measurement of mental effort seems to be promising, as it presumably allows
assessing subtle changes with respect to understandability~\cite{Zug+12b}. To
this end, we computed the average mental effort for each question.
Table~\ref{tab:mentaleffort} shows the mental effort for the specific questions
per model mentioned in Sect.~\ref{sec:methodology} (11 questions per model).

\vspace{-0.3cm}
\begin{table}[htb]
        \centering
        \begin{tabular}{lccc}
        \toprule
        &\textbf{Minimum}&\textbf{Maximum}&\textbf{Mean}\\
        \midrule
        $P_1$ graphical representation & 3.09 & 4    & 3.68 \\
        $P_1$ textual representation   & 3.36 & 6    & 4.47 \\\midrule
        $P_2$ graphical representation & 3.45 & 4.73 & 3.96 \\
        $P_2$ textual representation   & 4.27 & 4.82 & 4.48 \\
        \bottomrule
        \end{tabular} 
    \caption{Mental effort}
    \label{tab:mentaleffort}
    \vspace{-0.8cm}
\end{table}


\paragraph{Discussion.}
The empirical investigation indicates that answering questions to a graphically
represented model requires a higher mental effort than for a textual one.
To understand these findings, we would like to refer to the
\textit{Split--Attention Effect}~\cite{Kal+03}. This effect occurs when
information from different sources has to be integrated and is known to increase
mental effort. In our case, when studying a textually represented model that
consists of three separate lists (activities, constraints and an
activity--constraint mapping), the subject has to keep parts of one list in
working memory while searching for the matching parts in other lists. Thereby,
two basic effects are distinguished. First, the reader has to switch attention between
different information sources, e.g., constraint and mapping lists. Second, the
reader has to integrate different information sources. These two phenomena in
combination are then known to increase mental effort and are referred to as
split–-attention effect.


\subsubsection{$RQ_{2}$: What are advantages of each representation and what are
potential improvements for the understandability of declarative process models?}
\label{sec:rq2}
The goal of $RQ_{2}$ is to complement findings obtained so far with opinions
from students, i.e., subjective measures. In particular, after all specific
questions were answered, we additionally asked general questions for each model.
To analyze answers, we identified and classified issues, which---according to
the subjects---influence the sense--making of declarative business process
models. All in all, we could find 5 factors that subjects considered to be
harmful for the sense--making of declarative process models (cf.
Table~\ref{tab:ggG2}). Three subjects mentioned that the pairs
of constraints posed a considerable challenge for the sense--making
(cf.~\cite{BPMDS2013}). In addition, 3 subjects explicitly mentioned that they
experienced problems with the high number of constraints and resulting
dependencies (combination of constraints). One subject explained that he had
problems due to too many activities. Another one mentioned that he was
challenged making sense of $P_2$ because there were too many components. Also, 6
subjects perceived the textual representation as a negative influence.

\vspace{-0.3cm}
\begin{table}[htb]
    \centering
    \begin{tabular}{p{2cm}lcc}
    \toprule \textbf{Category} & \textbf{Factor} &
    \textbf{Subjects} & \textbf{Influence}\\ \midrule
    Constraints & Pairs of constraints & 3 & $-$ \\
                & Combination of constraints & 3 & $-$\\
                & Number of activities & 1 & $-$\\
                & Number of components & 1 & $-$\\
    \midrule
    Other & Representation & 5 & $-$\\ \bottomrule
    \end{tabular}
    \caption{Why do you think the model was (not) difficult to understand?}
    \label{tab:ggG2}
    \vspace{-0.6cm}
\end{table}

Regarding advantages or disadvantages of each representation, 6 subjects
mentioned that the graphical representation was easier to grasp. One
subject answered that the graphical representation is also unclear sometimes due to
pairs of constraints (cf.~\cite{BPMDS2013}). One subject praised the good
overview of the constraints at the textual representation. Overall, the subjects
had a better perception of the graphical representation, which might also be
concerned with the shorter duration and lower mental effort (cf. $RQ_{1.2}$ and
$RQ_{1.3}$).
Also, subjects made propositions how to make declarative process models easier
to understand. In particular, 7 subjects proposed to only use the graphical
representation. In addition, one subject indicated that paired constraints
should be simplified. Unsurprisingly, suggestions for the improvement of
declarative process models are closely connected to respective problems (cf.
Table~\ref{tab:ggG2}). In general, it can be observed that the basic building
blocks of declarative process models---activities and constraints---are rather
unproblematic. However, the combination of constraints and in particular pairs
of constraints, in turn, pose considerable challenges. In this sense, for
instance, approaches providing computer--based support for the interpretation of
constraints seem promising~\cite{Zug+11b}.

\subsubsection{Limitations.}
Our work has the following limitations. First, the number of subjects in the
empirical investigation is relatively low (9 subjects), hampering the only of
descriptive nature result's generalization. Second, even though process models
used in this investigation vary in the number of activities, number of
constraints and representation, it remains unclear whether results are
applicable to declarative process models in general, e.g., more complex models.
Third, all subjects are students, further limiting the generalization of
results. Finally, most errors were due to problems with the setup of the
investigation (cf. Table~\ref{tab:errorsQual}).

\vspace{-0.5cm}
\section{Related Work}
\label{sec:related_work}
In this work, we investigated the \textit{understanding} of graphical and
textual declarative process models. More generally, factors of conceptual model
comprehension were investigated in~\cite{MSR12}, and the understandability of
imperative process models was investigated in~\cite{ReMe}. Comparisons of
graphical and textual notations were examined from different angels. For
instance, the interpretation of business process descriptions in BPMN (graphical
notation) and in an alternative text notation (based on written use-cases) was
investigated in~\cite{OttensooserJSS2012}. More generally,~\cite{Whitley97}
provides an overview of relative strengthes and weaknesses of textual versus
flowchart notations.
For this investigation, we have focused on the declarative modeling language
Declare. Recently, also Dynamic Condition Response (DCR) graphs~\cite{HiMu10}
have gained increasing interest. Unlike Declare, DCR graphs focus on a set of
core constraints instead of allowing for the specification of arbitrary
constraints. However, so far, contributions related to DCR graphs have rather
focused on technical aspects, such as technical feasibility and expressiveness,
while understandability was not approached yet.



\vspace{-0.5cm}
\section{Summary and Outlook}
\label{sec:summary}
Declarative approaches to business process modeling have recently attracted
interest, as they provide a high degree of flexibility~\cite{Pesi08}. However,
the increase in flexibility comes at the cost of understandability and hence
might result in maintainability problems of respective process
models~\cite{Pesi08}. The presented empirical investigation presents differences
between graphical and textual represented declarative business process models.
The results indicate that the graphical representation is advantageous in terms
of errors, duration and mental effort. In addition, subjects themselves
expressed that the graphical representation is easier to understand. As
indicated in \cite{BPMDS2013}, it might be recommendable to avoid representing
declarative models in a way similar to imperative models, especially when
semantic differ considerably (cf. Mindshift Learning theory~\cite{ArHa07}).
However, a pure textual representation does not seem to be the right solution.
To accomplish our goal of a better understandability of declarative process
models, further investigations are needed. Particularly, replications utilizing
an adapted hybrid representation seem to be appropriate means for additional
empirical tests.

\vspace{-0.4cm}

\bibliographystyle{splncs}
\bibliography{literature}

\end{document}